\documentclass[12pt,preprint]{aastex}
\usepackage{apjfonts}
\begin{document}
\slugcomment{\aj, accepted}
\newcommand{\LeeZinn}{\mathscr{L}}

\title{Light Curves and Period Changes of Type II Cepheids in the Globular Clusters
M3 and M5}

\author{Katie Rabidoux \altaffilmark{1} \altaffilmark{9}, Horace A. Smith \altaffilmark{1}, Barton J. Pritzl \altaffilmark{2} \altaffilmark{4},
Wayne Osborn \altaffilmark{3} \altaffilmark{8}, Charles Kuehn \altaffilmark{1}, Jill Randall \altaffilmark{1}, R. Lustig \altaffilmark{2}, K. Wells \altaffilmark{1}, Lisa Taylor \altaffilmark{1}, Nathan De Lee \altaffilmark{1} \altaffilmark{5}, K. Kinemuchi \altaffilmark{1} \altaffilmark{5}
\altaffilmark{6}, 
Aaron LaCluyz{\'e} \altaffilmark{1}, D. Hartley \altaffilmark{1}, C. Greenwood \altaffilmark{1},  
M. Ingber \altaffilmark{1}, M. Ireland~\altaffilmark{1},  E. Pellegrini
\altaffilmark{1}, Mary Anderson \altaffilmark{1}, Gene Purdum \altaffilmark{1}, J. Lacy \altaffilmark{3}, M. Curtis
\altaffilmark{3}, Jason Smolinski
\altaffilmark{1} \altaffilmark{3}, and Stephen Danford \altaffilmark{7}}

\altaffiltext{1}{Department of Physics and Astronomy, Michigan State University, East Lansing, MI 48824}
\altaffiltext{2}{Department of Physics and Astronomy, Macalester College, 1600 Grand Avenue, Saint Paul, MN 55105}
\altaffiltext{3}{Department of Physics, Central Michigan University, Mount Pleasant, MI 48859}
\altaffiltext{4}{Current Address: Department of Physics and Astronomy, University of Wisconsin Oshkosh, 800 Algoma Blvd., Oshkosh, WI, 54901}
\altaffiltext{5}{Current Address: Department of Astronomy, University of Florida, 211 Bryant Space Science Center, Gainesville,
FL 32611} 
\altaffiltext{6}{Departamento de Astronom\'{i}a, Universidad de Concepci\'{o}n, Casilla 160-C, Concepci\'{o}n, Chile}
\altaffiltext{7}{Department of Physics and Astronomy, University of North Carolina-Greensboro, PO Box 26170, Greensboro, NC 27402}
\altaffiltext{8}{Yerkes Observatory, 373 West Geneva Street, Williams Bay, WI 53191}
\altaffiltext{9}{Current Address: Department of Physics, Hodges Hall, Box 6315, West Virginia University, Morgantown, WV 26506-6315}

\begin{abstract}

Light curves in the $B$, $V$, and $I_c$ passbands have been obtained for the type II Cepheids
V154 in M3 and V42 and V84 in M5. Alternating cycle behavior, similar to that seen among RV
Tauri variables, is confirmed for V84.  Old and new observations, spanning more than a century,
show that V154 has increased in period while V42 has decreased in period.  V84, on the other hand,
has shown large, erratic changes in period that do not appear to reflect the long term evolution 
of V84 through the HR diagram.

\end{abstract}

\keywords{stars: variables: Cepheids --- globular clusters: individual (NGC 5272, NGC 5904)}

\section{Introduction} 
Type II Cepheids were among the first variable stars discovered within the
globular clusters M3 (NGC 5272) and M5 (NGC 5904) \citep{p89, p90}. Their initial
discovery came long before the realization that Cepheids were pulsating
stars \citep{sh14}, and even longer before Baade's \citep{b56} discovery that Cepheids
could be divided into two population groups. Until recently,
the vast majority of observations available
for globular cluster Cepheids were obtained photographically.  This paper
presents new $B$, $V$, and Cousins $I_c$ band
CCD light curves of the type II Cepheids V154 in M3 and V42 and V84 in
M5, as well as some new photographic data.  These variable stars were first observed more than a century
ago, raising the possibility that their long term period changes may indicate
the direction and speed of their evolution through the instability strip.
Our new light curves, together with earlier observations and, in the case of V42,
observations from the All Sky Automated Survey (ASAS) \citep{po02}, are
used to rediscuss the long term period changes of these variables.

\section{Observations and Reductions} 

\subsection{New CCD Data}

Images of M3 and M5 covering 10 x 10 arcmin were taken with the 0.6-m reflector on the campus of
Michigan State University during 2003-2005 using an Apogee Ap47p CCD
camera. Additional Michigan State University observations were obtained of M5 in 2006
utilizing an Apogee Alta U47 CCD camera. Supplemental images were also obtained with 
two 0.4-m reflectors - in 2003 at the Brooks Astronomical 
Observatory of Central Michigan University with a Photometrics Star-1 camera and in 2004 with an 
SBIG ST-8 CCD on the Macalester College telescope. 

As expected for observations from low altitude midwestern sites, seeing was usually not good,
often being around 3 seconds of arc and sometimes worse. Images were obtained with Johnson
$B$, $V$, and Cousins $I_c$ filters. Exposure times ranged from one minute to six minutes 
depending on telescope and filter.  Bias, dark, and flat field corrections
were applied using standard techniques. The data were then reduced using Peter 
Stetson's DAOPHOT profile-fitting reduction package \citep{st87,st94}. The resulting instrumental 
b, v and i magnitudes were reduced to the standard system using equations of the form:

\begin{equation}
 V = v + {a_1}(b-v) + {c_1}; 
\end{equation}

\begin{equation}
B = b + {a_2}(b-v) + {c_2};
\end{equation}

\begin{equation}
I_c = i + {a_3}(v-i) + {c_3}.  
\end{equation}

Color terms were determined from observations of Landolt standard stars and stars
within the open cluster M67 \citep{sc83, sc85, la92}.
Occasionally, observations were not made through all three filters so that the usual means 
of obtaining color corrections could not be applied. In those cases, the color at the time of 
observation was determined from the typical color at the appropriate phase in the light curve. The color 
terms are sufficiently small and the light curves are sufficiently well established that this is 
not expected to be an important source of uncertainty. 

Zero-point corrections to the standard system
were determined from 5-6 uncrowded local standard stars within each cluster for 
which magnitudes
had been determined by \citet{st00}. The type II Cepheids tend to be bluer than
the other bright, uncrowded stars within the fields of our CCD images, and thus the
local Stetson standards are redder (by 0.5 to 0.8 in $B-V$) than the target variables in both M3 and M5.
The redder colors of the local standards will introduce error in the standard magnitudes 
for the Cepheids if there is a significant error in the coefficients of the color transformation 
equations. We have avoided extremely red local standards, and the transformation 
coefficients were determined separately for the Michigan State University, Central 
Michigan University, and Macalester College observations, which should minimize 
uncertainties from this source. Nonetheless, systematic errors of about 0.01 magnitude are 
possible. 

The photometry of the three Cepheids is listed in Table 1. The filter, variable star 
identification, heliocentric Julian date, observed magnitude, uncertainty, and the source
of the individual observations are 
listed in columns 1 through 6, respectively. In column 6, MSU is used for observations
obtained at the Michigan State University observatory, CMU for observations obtained at the
Central Michigan University observatory, and MAC for observations obtained at the observatory
of Macalester College. The uncertainties listed in Table 1 reflect the formal uncertainties in the photometry as 
propagated through the daophot program and the application of the transformation 
equations. 
Heliocentric Julian Dates were determined from UTC of mid-exposure without application 
of the, in these cases, insignificant $\Delta$T correction for the difference between UTC and 
uniform ephemeris time \citep{st05}.

\subsection{New Photographic Data}

Because relatively few observations of V42 and V84 in M5 had been published between 1976
and 2003, one of us (W.O.) searched the plate vault of the Yerkes Observatory for plates of
this cluster taken during this time period. Fourteen useful plates, all taken in 1982-83, were located. 
These are IIa-O plates taken with a BG1 filter, so that magnitudes estimated from these plates
approximate those in the B photometric bandpass. W.O. made eye estimates of V42 and
V84 on the plates, interpolating among standard stars A, B, E, and F in \citet{cc77}, adopting the 
B magnitudes for these stars given in Arp (1962).
The estimated uncertainty of each estimate is about 0.15 magnitude. The resultant
photometry is listed in Table 2.

\subsection{ASAS Data}
V42 is sufficiently far from the center of M5 that V-band photometry for it has been
obtained in ASAS (Pojmanski 2002). We have extracted 
353 $V$-band observations labeled of quality A (the best quality) from the photometry list
for the ASAS object designated 151825+0202.9. Because the pixels of ASAS are relatively big
(15 arc seconds), we have used only data from the smallest
ASAS aperture (2 pixels in diameter). The listed uncertainties on the magnitudes are
typically 0.03 to 0.05 magnitudes.

\section{Periods and Light Curves}

The best periods for the variables were determined using the {\rm Phase Dispersion Minimization} routine \citep{st78} as
implemented in IRAF \footnote{IRAF is distributed by the National Optical Astronomy Observatories,
    which are operated by the Association of Universities for Research
    in Astronomy, Inc., under cooperative agreement with the National
    Science Foundation.}, the {\rm period04} program \citep{l05}, and a discrete
Fourier transform as implemented in the {\rm Peranso} 
software suite \footnote{http://www.peranso.com/}.  These methods all yielded
consistent results and provided periods
which were then used to construct phased light curves. 
The results for each star are presented below. The adopted uncertainties in the derived 
periods are a combination of the uncertainties given by the {\rm period04} routine and 
from visual inspection of the light curves.

\subsection{V154 in M3}

The type II Cepheid V154 in M3, discovered in 1889, was the first periodic variable star to be identified within a
globular cluster \citep{b90}.  However, because of its proximity to the center of the 
cluster, it has been omitted in many photometric studies of the variable stars in M3. 
The best period for our 2003-2004 data was found to be 15.29 $\pm$ 0.02 days, close to the
period of 15.2842 days used by \citet{a55} and adopted in the O-C diagram of \citet{h80}.  The 
$B$, $V$, and {$I_c$ phased light curves of V154 in M3 are shown in Figure 1.

The scatter about the mean light curves in Figure 1 is larger
than the formal photometric uncertainties (typically 0.02 or 0.03 mag). V154 is 
close enough to the center of the cluster that there can be some blending with neighboring
stars, especially on our nights of poorer seeing, and that is very likely responsible for some
of the scatter in the light curves despite our use of a profile-fitting photometry technique.  
\citet{ba00} note that in their observations the image of V154
is blended with that of an RR Lyrae star, V268, which would also be the case with our
observations. Nonetheless, it is possible that some of the scatter reflects real changes in the 
variability of V154.  In a study of the light curve of the field type II Cepheid
W Virginis (main period = 17.27 days), 
\citet{th07} noted that the light curve of that star showed a scatter of about 0.1 magnitude, much 
larger than the 0.01 mag expected from observational error alone. They concluded that the 
light curve of W Vir could not be completely described by a single periodicity, and were able to 
identify two additional periodicities that contributed to the observed light curve. Our data for 
V154 are less extensive than the W Vir data obtained by \citet{th07}, and are 
not adequate to reveal any secondary periodicity for V154 comparable to those found in W Vir. 
However, the possibility that some of the scatter in our light curves of V154 may reflect 
real cycle to cycle differences should be kept in mind. 

\subsection{V42 in M5}

In contrast to V154, V42 in M5 is relatively uncrowded. The derived period for V42 from our 
2003-2006 observations is 25.735 $\pm$ 0.015 days, and this has been used to produce the 
light curves shown in Figure 2. The scatter in the light curve of V42, though smaller than that of V154, is still slightly 
larger than expected from the formal observational errors. Though some of this may 
reflect sources of observational uncertainty not included in the formal error analysis, we 
cannot exclude real fluctuations in the light curve as a source of scatter. 
For comparison, the period given by \citet{cc77} is 25.738 days.  The maxima, minima, and mean 
magnitudes derived from our $B$ and $V$ light 
curves are in good agreement with those from the more sparsely covered $B$ and $V$ light curves 
observed photoelectrically by \citet{a57}.

The light curves of V42 from the Yerkes and ASAS data are shown in Figures 3 and 4, respectively.  The Yerkes data are 
plotted using the same 25.735 d period adopted for the CCD light curves. While the points 
are sparse and have large uncertainties, the maximum and minimum are consistent with the 
CCD photometry for B.  The period that 
best fits the ASAS observations from JD 2451930 until 2455057 (2001-2009) is 25.720 $\pm$ 0.003 days, 
and this has been used to construct the ASAS light curve.  The V amplitude from 
the ASAS data is smaller than seen in our observations or in those of Arp (1957), and the mean 
magnitude is brighter. This probably occurs because, even with the smallest aperture, the 
ASAS pixels are too big to eliminate the contribution of neighboring stars to the
aperture photometry.  Nonetheless, the large number of data points and the long
interval of time coverage make the ASAS observations of V42 very useful for the discussion
of its period changes.

\subsection{V84 in M5}

V84 is in a more crowded field than V42, and its photometry undoubtedly suffers from that 
circumstance.  The period determination for V84 also turned out to be more complicated than for V42. 
We first consider only our observations from 2003 through 2005. The V84 photometry for these years 
can be approximately fit with a period of 26.93 $\pm$ 0.02 days.  Figures 5 and 6 show the light curves 
for this period for the CCD and Yerkes data, respectively.  This period, however, leaves larger than 
expected scatter in the CCD light curves. It is also significantly longer 
than the period of 26.42 days used by \citet{cc77}.

\citet{a55} suggested that, while a period of 26.5 days described the main pulsation of V84, the 
light curve of V84 might be better described with
a period twice as long. We carried out a period search in the vicinity of twice 26.9 days and
obtained a best fit with a period of 53.95 $\pm$ 0.03 days.
Figure 7 shows the 2003-2005 light curve of V84 plotted with this period, which is approximately but not
exactly twice 26.93 (53.86 d).  The scatter in the observed curves is reduced, although there
are some gaps in phase coverage. The light curves show some evidence of alternating deep and
shallow minima. RV Tauri stars are known to sometimes show alternating deep and shallow minima,
e.g., \citet{g92}, and the light curve of V84 might therefore be indicative of low-level 
RV Tauri type behavior. 

More complications arise when the observations from 2006 are included. The 2006 observations cannot be
well-phased with the earlier data with either the 26.93 day or the 53.95 day periods, in both cases 
showing a significant phase shift between the two data sets (see Figure 8). In fact, no single period can 
provide a light curve for V84 that does not show large scatter when observations from 2003 through 2006 
are combined. As we discuss below, abrupt changes in the period, and hence phase shifts, of V84 have 
been noted before, and the 2006 observations likely indicate another such jump.

While the period solutions for V42 in M5 have shown little change over time, that is 
not the case for V84.  This will be discussed in detail in \S 4.2, but we note that 
our period of 26.93 days is significantly longer than the 26.42 days used by \citet{cc77} in plotting their
light curve. In any particular observing
season, lasting 3 or 4 months, the difference between a light curve with a 26.42 day period
and a 26.93 day period is not large (less than about 0.05 in phase).  However, over the
span of two observing seasons, i.e. about a year, the difference can amount to a fifth of a cycle.   
The smaller number of observations in 2006 makes it impossible to determine the exact period 
during that year, and we cannot distinguish between
a period of 26.4 and 26.9 days from the 2006 data alone.

\section{Period Changes}

\subsection{M3}

The long term period behavior of V154 in M3 was studied by \citet{h80}, using observations made
between Julian dates 2416604 and 2442862, a span of 72 years from 1904 to 1976. We have expanded 
the interval of the period study to
just over a century.  In Table 3, we list the heliocentric Julian date (HJD) representing the epoch 
of maximum as determined
from a given set of observations, the phase of
maximum light calculated from the ephemeris of \citet{a55} ($JD_{max} = 2424627.55d + 15.2854 E$,
where $JD_{max}$ is the date of maximum light and E the cycle number),
the estimated uncertainty of the phase of maximum, and the source of the data.  We have combined the
three closely spaced maxima reported in \citet{a55} into a single representative point.  We have chosen 
to be conservative in assessing the accuracy of the times of maximum, and thus in some instances assign 
larger uncertainties than were used by \citet{h80}. Unpublished observations of V154
from the study of M3 variables by \citet{st02} were used to add an additional epoch of maximum.

Figure 9 shows a possible increase in period for V154 early in the observational record and, with
more certainty, an
increase in period after JD 2,450,000.  A period of 15.2854 days adequately represents the observations
made between JD 2,420,000 and JD 2,450,000.  The sudden increase in phase for the more recent observations
indicates an increase in period, but the gap in the observations before JD 2451256 makes it difficult
to tell exactly when that period increase happened.  If we assume an abrupt increase in period near
JD 2,450,000, then we find that the period increased to about 15.296 days. That period is larger than
the value of 15.29 $\pm$ 0.02 days found from the 2003-2005 observations alone, but it is within the
estimated one sigma error bar of our period.

\subsection{M5}

\citet{cc77} studied the long term period behavior of V42 and V84
using mainly photographic data spanning the 87-year interval from 1889 to 1976.  To their compilation of
data we can add the results from our observations, plus several additional results from
observations made by others since 1976, which extend the studied time interval to 
about 120 years for V42 and 109 years for V84.

In discussing the period changes of V42, we have adopted the fiducial period 
(25.738 days) and
epoch of zero-phase (HJD 2441102.7) used in \citet{cc77}.  Additional epochs of maximum for V42 not included in
\citet{cc77} are listed in Table 4 which includes an epoch of maximum determined from 
unpublished photometry of V42 provided by T.~M. Corwin (private comm.).  \citet{cc77} concluded that 
the period of V42 had been relatively stable
with a period near 25.738 days since 1889, but that there had been a small period decrease of about
0.007 day.  They noted that the change could have been occurring continuously, or that
there could have been an abrupt period decrease in the 1940s.  In the phase diagram for V42
shown in Figure 10, a more dramatic decrease in period is apparent.  Until JD 2,435,000, the
phase diagram is well described by a period of 25.738 $\pm$ 0.004 days.  As found by \citet{cc77}, there
is evidence for a slight decrease in period after that date. Between JD 2,435,000
and JD 2,441,000 the phase diagram can be well fit with a period of 25.731 $\pm$ 0.004 days.  Between
JD 2,441,000 and JD 2,455,000, the phase diagram is well fit with a period of 25.720 $\pm$ 0.003 days.
As in the study of \citet{cc77}, the phase diagram does not let us determine whether the
period changes are actually abrupt.  However, the phase diagram in Figure 10 is slightly better fit
by three straight line segments than by the parabola that would indicate a constant
rate of period change. \citet{b95} in their abstract reported a period of 25.725~days based upon
their unpublished photometry of V42, consistent with the decline in period found here.

In addressing the period change behavior of V84, we adopted a fiducial period
of 26.42 days and epoch of zero-phase of HJD 2441129.6, again consistent with those
used in \citet{cc77}.  Additional epochs and phase shifts, 
beyond those given in Table V of \citet{cc77} are listed in Table 5.  Following \citet{cc77},
we use only the shorter period for V84 and not the 53 day double period in discussing the
period change behavior.  The observations in the literature are often not adequate for
addressing phase shifts using the longer period, but its neglect may introduce some
extra scatter into the phase shift diagram. The shifts in
phase versus Julian Date are shown in Figure 12.  In that figure we are faced with a much
more confusing situation than was evident in Figures 9 or 10, a circumstance to which
\citet{cc77} have already called attention.  The scatter in the phase shifts implies changes
in period (or jumps in phase), and the changes are sufficiently large that it is not always clear whether
the count of cycles between observed epochs is correct.

\citet{b1898} determined the period of V84 to be 26.2 days based upon
visual observations obtained with the Yerkes 1-m telescope.
\citet{a55} obtained photographic observations of the M5
Cepheids and later reported additional photoelectric observations for the pair \citep{a57},
confirming his earlier report of the existence of alternating cycle behavior for V84.  The periods 
given in \citet{a57}
are 26.62 $\pm$ 0.03 days for the shorter cycle, and 53.24 $\pm$ 0.2 days for the doubled cycle. 
\citet{w58} used his observations and those of \citet{a55} to determine a period of 26.54 days but also 
found evidence for alternating minima.   By far
the most extensive previous study of the period of V84 is the study of \citet{cc77}. They determined 
that during the 1930s and 1940s, the period of V84 remained nearly
constant at 26.42 days.  They found that during the 1950s, the period increased by about 0.2 days
before decreasing again.  There may have been another period jump in 1970, but by 1971 the
period had settled again at 26.42 days.  Our light curves of V84, and Figure 11, indicate an
even more extreme increase in period to 26.93 days between 2003 and 2005.  The phase shift shown 
in Figure 8 between the 2003-2005 and 2006 light curves for V84 suggests that its period has declined again, 
perhaps to near 26.8 days.  Further observations are needed to confirm the exact value of the period
for 2006 and later years.

\section{Location in the Color-Magnitude Diagram}

Long period type II Cepheids such as V154 and V42 (and also V84, if one
includes variables with stronger RV Tauri-like behavior) tend to be brighter
than expected from an extrapolation of the period-luminosity
relation as determined from shorter period type II Cepheids (see, for example,
Figure 5 in \citet{bo97}). In order to place V154, V42, and V84 onto a color-magnitude diagram, 
we derive their intensity-weighted mean $V$ magnitude and
magnitude-weighted colors, $(B-V)$ and $(V-I)$.  These values and their
uncertainties are listed in Table 6. Absolute $V$ magnitudes,
as derived below, are also listed. The estimated uncertainty
for $<V>_{int}$ is about 0.02 for V42 and 0.03 for V84 and V154.
The estimated uncertainties for the mean colors are about 0.03 for
V42 and 0.04 for V84 and V154.

\citet{a57} derived a mean $V$ magnitude of 11.22 for V42 with a mean
$B-V$ color of 0.60, in good agreement with our results.  \citet{a57}
did not obtain a large enough number of observations to derive mean
magnitudes and colors for V84. \citet{car98} refer to unpublished
$BV$ photometry of V42 which gave $<V>$ = 11.15, 
which is slightly brighter
than our value. We have not found in the literature any complete light curves of V154
on the $BVI_c$ system, but partial light curves are given in
\citet{be06}. It is not possible to calculate mean magnitudes from the
\citet{be06} observations, and their observations show significant
scatter at a given phase (as do ours), but their magnitudes may be slightly
brighter than ours.

To determine absolute $V$ magnitudes ($M_V$) for the Cepheids, we referenced
their brightnesses to the RR Lyrae variables in M3 and M5.  The mean
$<V>_{int}$ magnitudes for RR Lyrae stars in M3 and M5 are about 15.64 and
15.07, respectively \citep{ca05,r96,st01}.  We convert these to absolute
magnitudes using an absolute magnitude of 0.59 for RR Lyrae stars
of [Fe/H] = -1.5 and $\Delta M_V/\Delta [Fe/H] = 0.214 $ \citep{ca03}.
Assuming [Fe/H] = -1.5 for M3 and -1.2 for M5, \citep{zw84,ca05,y08}, we
obtain the absolute magnitudes shown in Table 6.
The resultant locations of the variables in the color-magnitude diagrams are
shown in Figure 13, adopting $E(B-V)$ = 0.01 for M3 and 0.03 for M5
\citep{ca05,r96,st01}. Also plotted in Figure 12 are type II Cepheids in
globular clusters from the tabulation in \citet{n94}. Following
Table 4 in \citet{n94}, a few variables with slightly discordant measures
by different observers are plotted more than once. All three of our
variables, but especially V42 and V84, fall near the upper bound of the
type II Cepheid instability strip, and near the transition to RV Tauri behavior
(see also \citet{wc84} and \citet{bo97}).

\section{Discussion of the Period Changes}

Theory predicts that long period type II Cepheids enter the instability strip either
while undergoing blueward instability loops from the asymptotic red giant branch as
a consequence of helium shell flashes, 
or during final blueward evolution as the hydrogen burning shell nears the
surface of the star \citep{sch70, m73, g85, cl88, bo97}. \citet{bo97} found that,
for the lower mass but brighter Cepheids, the instability strip could be
crossed two or three times as a consequence of thermal pulses.

The period of a pulsating star is linked to its density via Ritter's
pulsation equation, $Q = P \sqrt{\rho}$, where Q is the pulsation
constant, P is the period, and $\rho$ is the mean stellar density.  The pulsation period
of a Cepheid is often its most accurately known property, and, as noted long
ago by \citet{ed18}, a small change in the structure of the Cepheid will
reveal itself as a change in pulsation period before it can be
recognized in any other measured quantity. Each of our three stars showed
long term period change behavior, but the changes are different in each
case.  V154 showed a modest increase in
period consistent with movement to the red in the instability strip.
V42 showed a decrease in period, consistent with movement to the blue.
If these period changes indicate the long term evolution of these
stars, V154 could be interpreted as being on the redward evolving, and V42
on the blueward evolving portion of the instability loops predicted by
theory during shell helium burning. Alternatively, blueward
moving V42 might be in the final blueward evolutionary phase.
In neither
case, however, is a parabolic fit to the phase diagram,
implying a constant rate of period change, significantly better
than the assumption
of abrupt period changes. Using the theoretical timescales of \citet{g76},
\citet{cl88} found that one might expect a rate of period decrease of 
$P^{-1}dP/dt = -0.0005$ to $-0.002$ cycles per 100 years during the final
quiescent blueward evolving stage. The observed
decrease in the period of V42 is about $\Delta P/P = -0.0007$ over 120 years,
consistent with that expectation.

Although V42 and V84 are both near the luminosity dividing type II Cepheid and
RV Tauri behavior in the HR diagram \citep{wc84,bo97}, 
V84 showed period changes much more erratic than those of V42.
V84 is not, however, the only type II Cepheid exhibiting period
fluctuations. \citet{cl88} found that
period fluctuations were not unusual in type II Cepheids in globular clusters,
although rarely do they seem to reach the extent exhibited by V84. Apparently
random period fluctuations have also been observed in the O-C diagrams of other
variable stars \citep{be09, tu09}.
V1, 
a 15.5~day period Cepheid in the globular cluster M12, does perhaps show jumps
in the phase shift diagram on a scale similar to that of V84 \citep{cl88}.
V84 shows evidence of RV Tauri behavior, and strong cycle-to-cycle period 
fluctuations have been observed for RV Tauri
stars in the field, e.g. \citet{p05}.  However, \citet{cl88} do not report RV Tauri
type behavior for V1 in M12, so that these irregular periods appear not to be
limited only to long period variables near the RV Tauri domain in the HR diagram.

\section{Acknowledgements}

We thank the National Science Foundation for partial support of this work under
grants AST0440061, AST0607249, AST0707756, and PHY0754541,
and the Yerkes Observatory for granting access to their extensive archive 
of photographic plates.  We thank Mike Corwin for providing unpublished
observations of V154 and V42, respectively. We thank Christine Clement for helpful comments on a
draft of this paper.

\begin{figure}
  \includegraphics{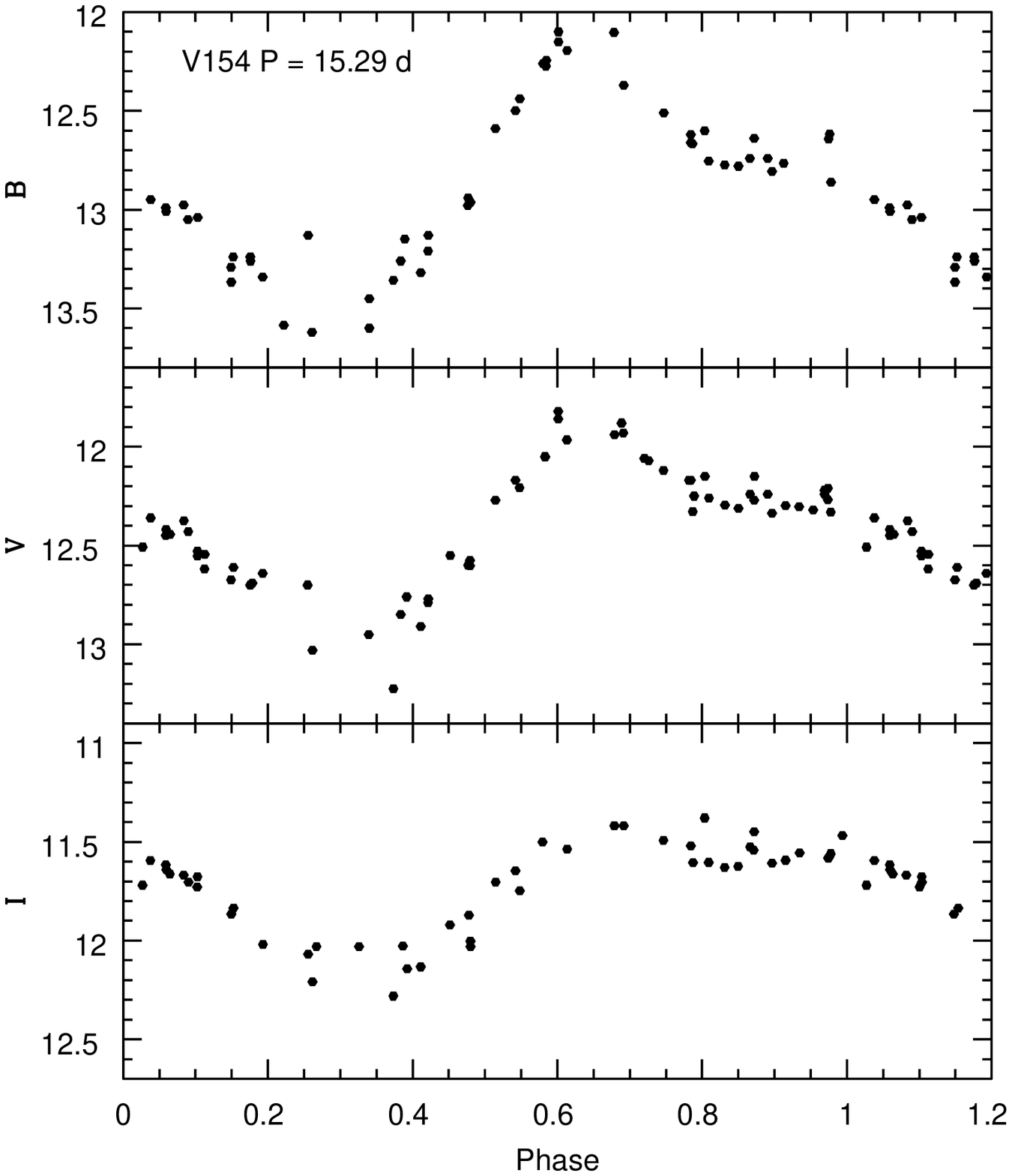}
  \caption{Light curves for V154 in M3, phased with a period of 15.29 days. The scatter
in the light curve is larger than the formal photometric uncertainties, typically 0.02
or 0.03 magnitudes.
      }
\end{figure}

\begin{figure}
\includegraphics{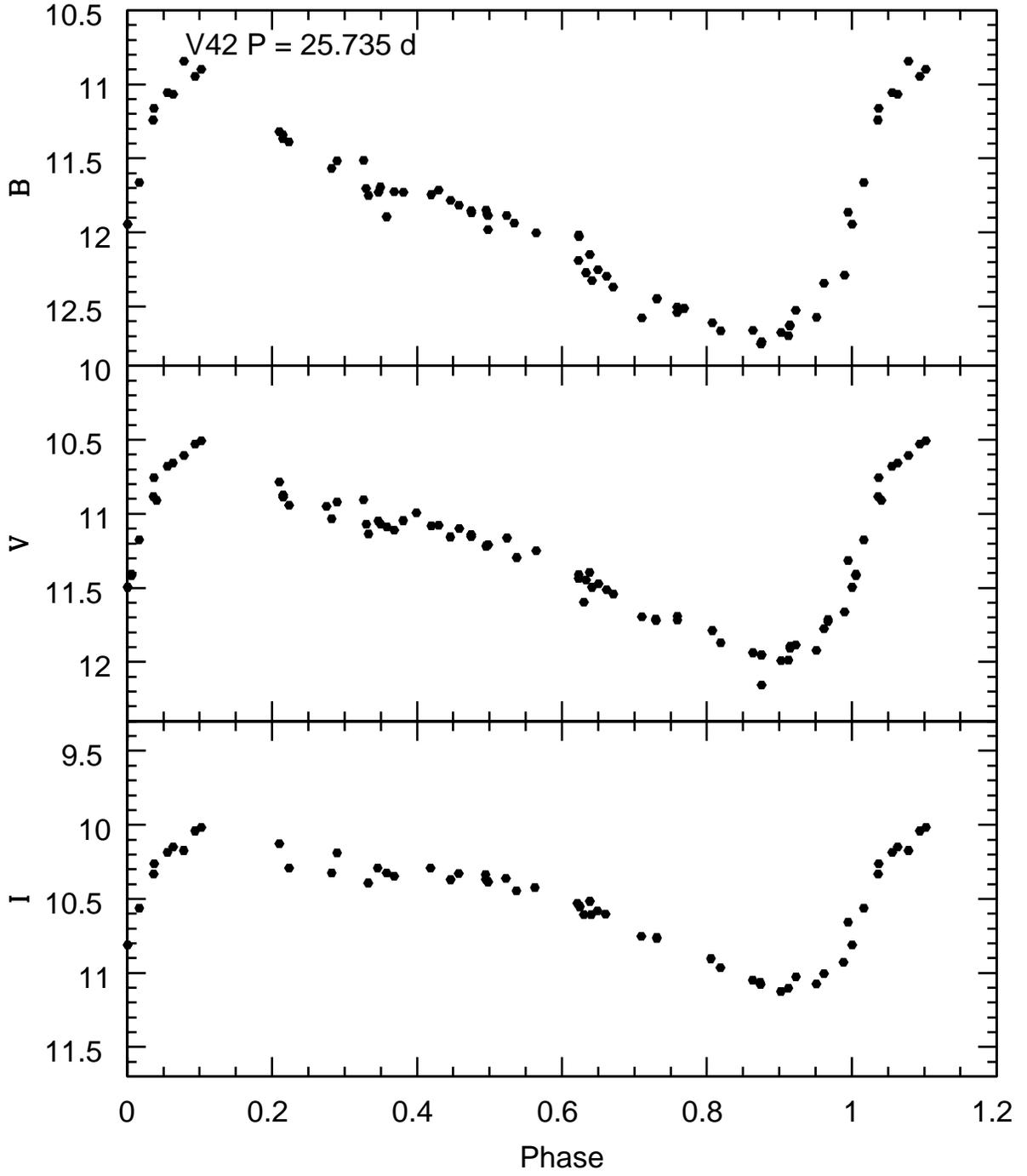}
\caption{Light curves for V42 in M5, phased with a period of 25.735 days.}
\end{figure}

\begin{figure}
\includegraphics{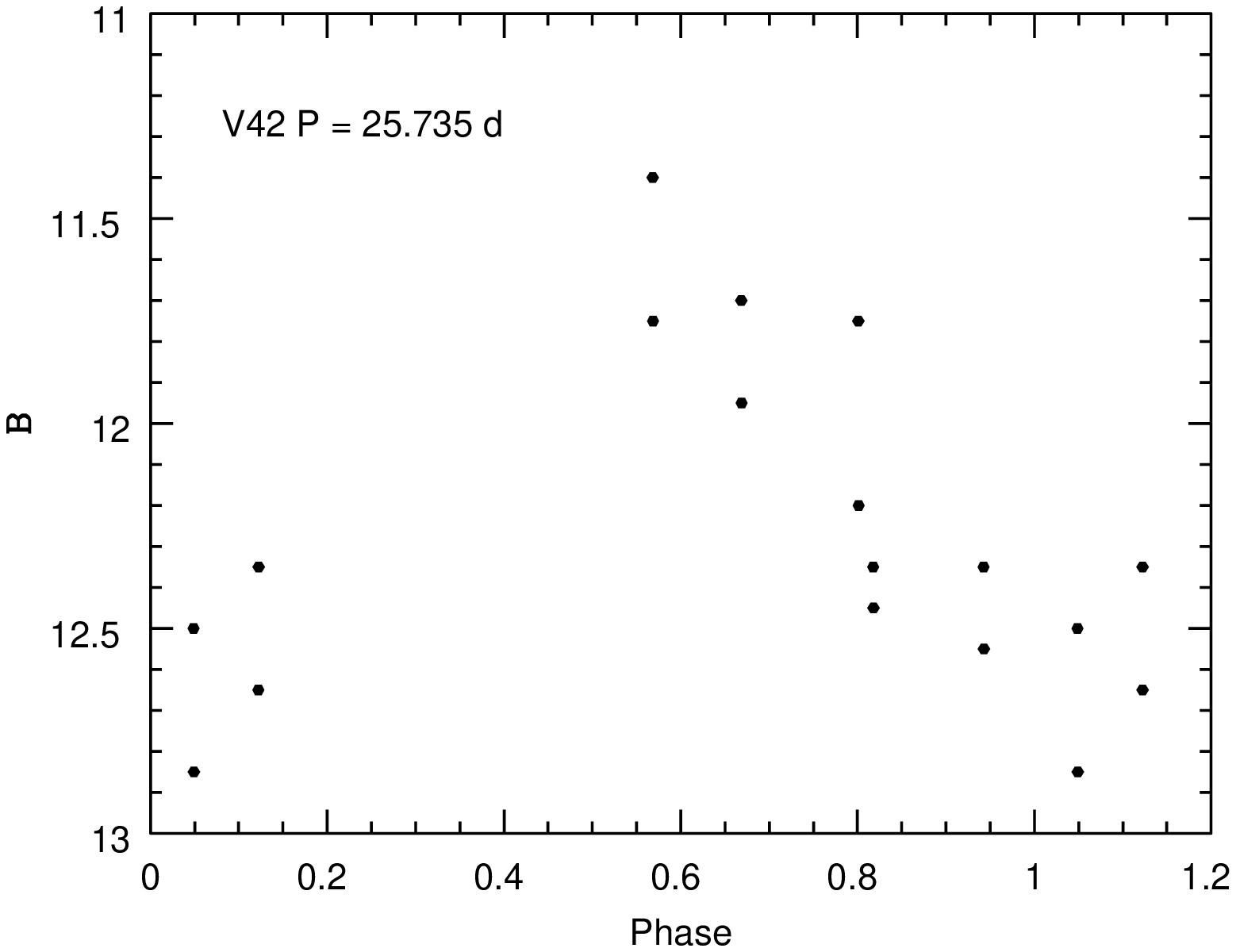}
\caption{$B$ band light curve of V42 based upon
Yerkes photographic observations, phased with a period of 25.735 days}
\end{figure}

\begin{figure}
\includegraphics{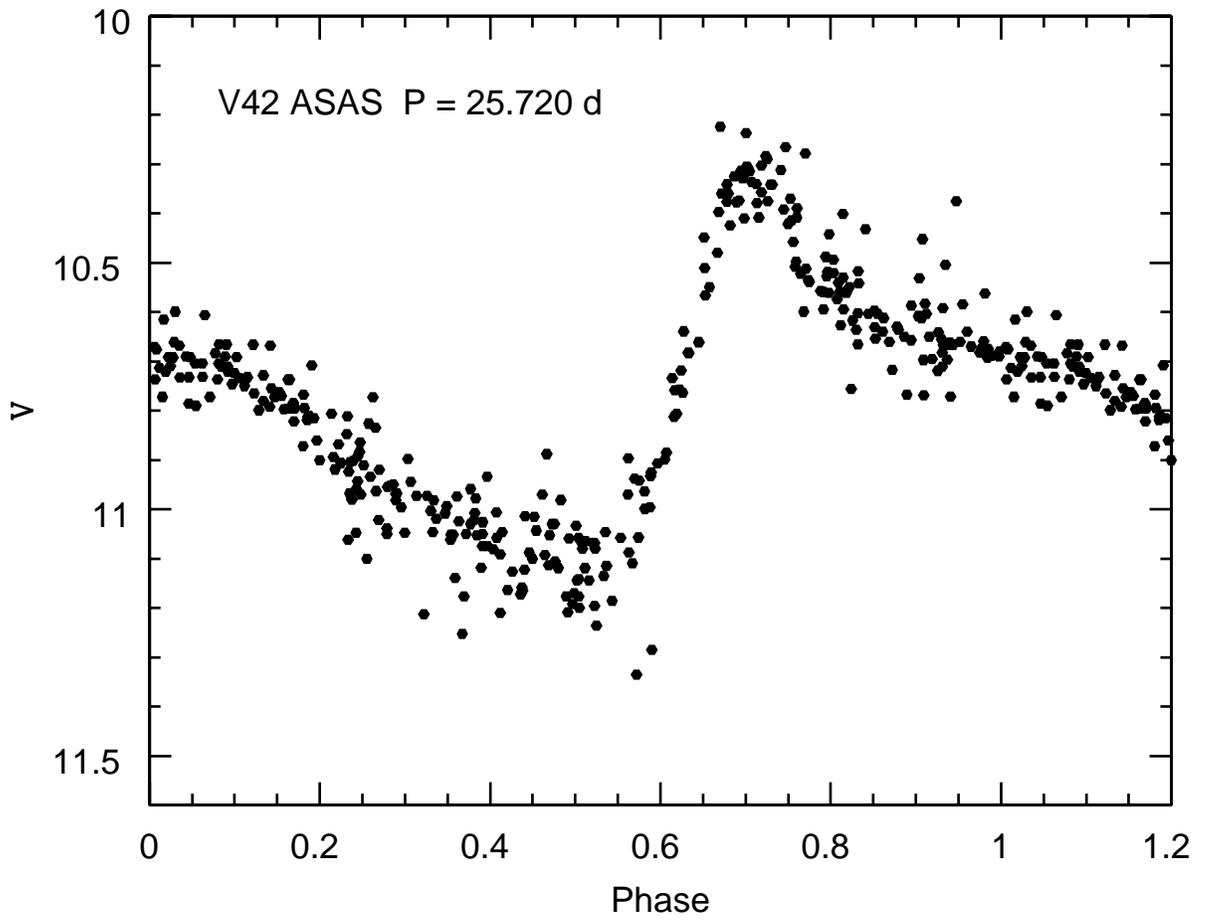}
\caption{Aperture photometry of V42 using ASAS data, phased with a period
of 25.720 days.
      }
\end{figure}

\begin{figure}
\includegraphics{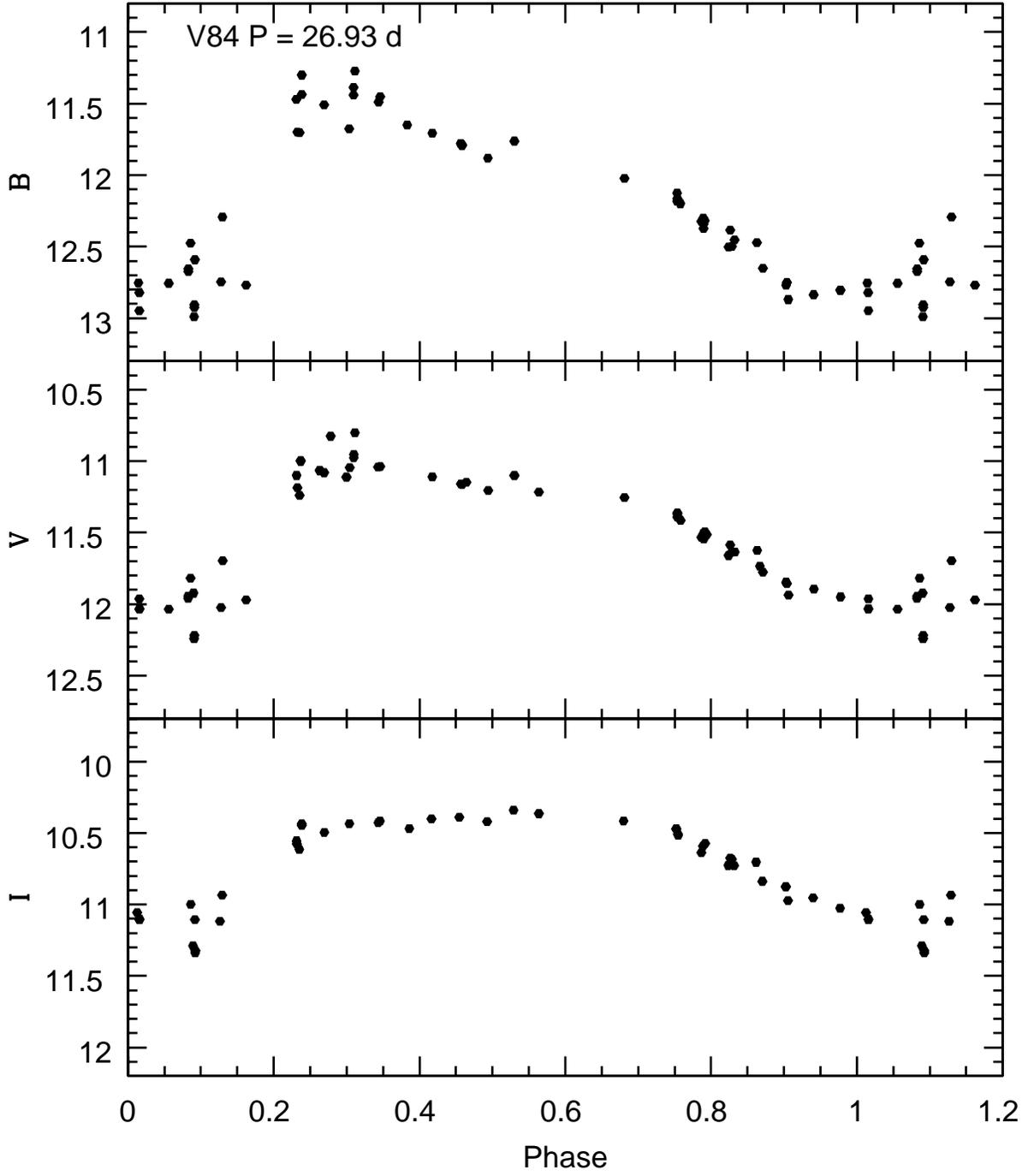}
\caption{CCD light curves for V84 in M5, phased with a period of 26.93 days.
      }
\end{figure}

\begin{figure}
\includegraphics{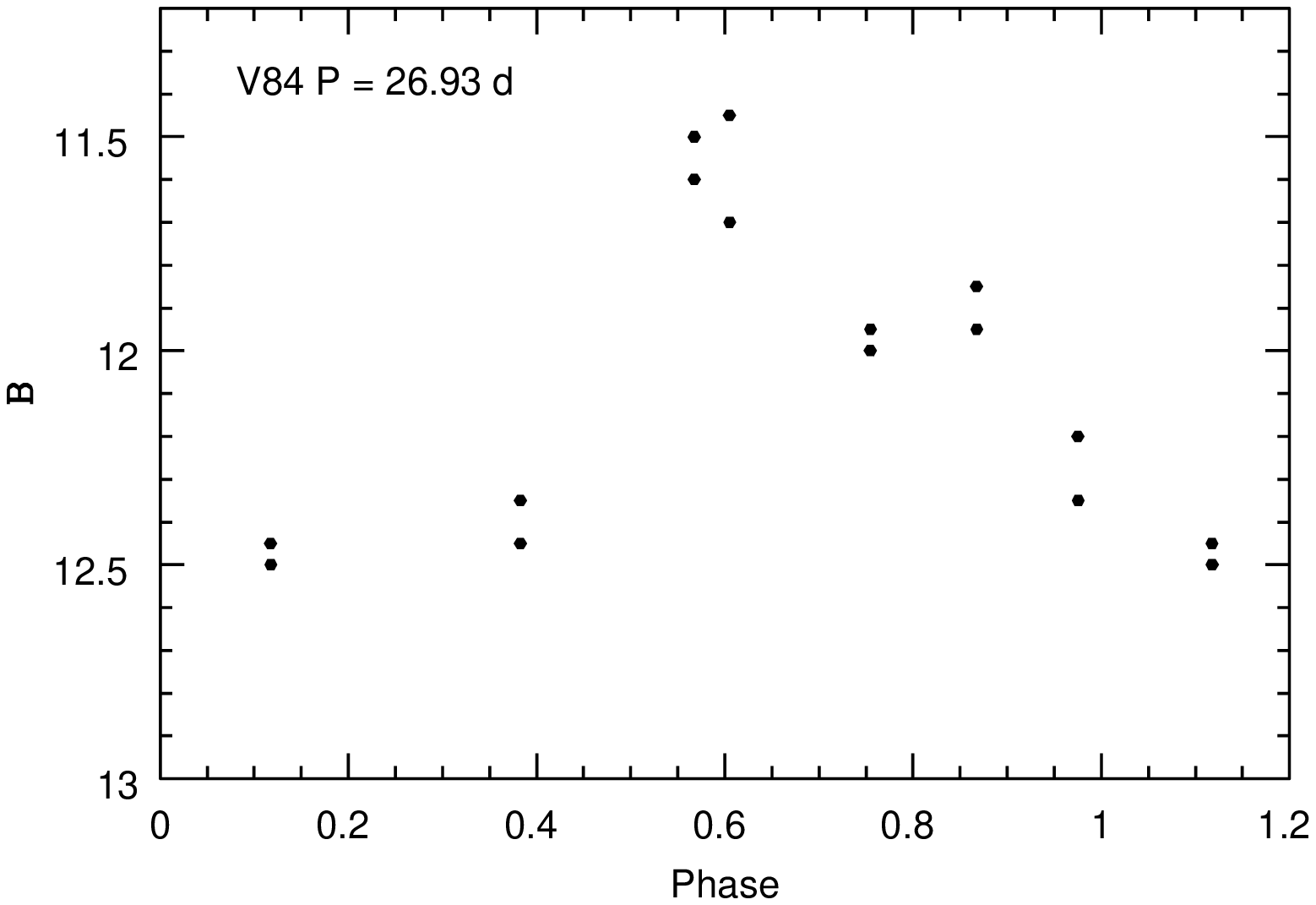}
\caption{$B$ band light curve of V84 based upon
Yerkes photographic observations, phased with a period of 26.93 days.
      }
\end{figure}

\begin{figure}
\includegraphics{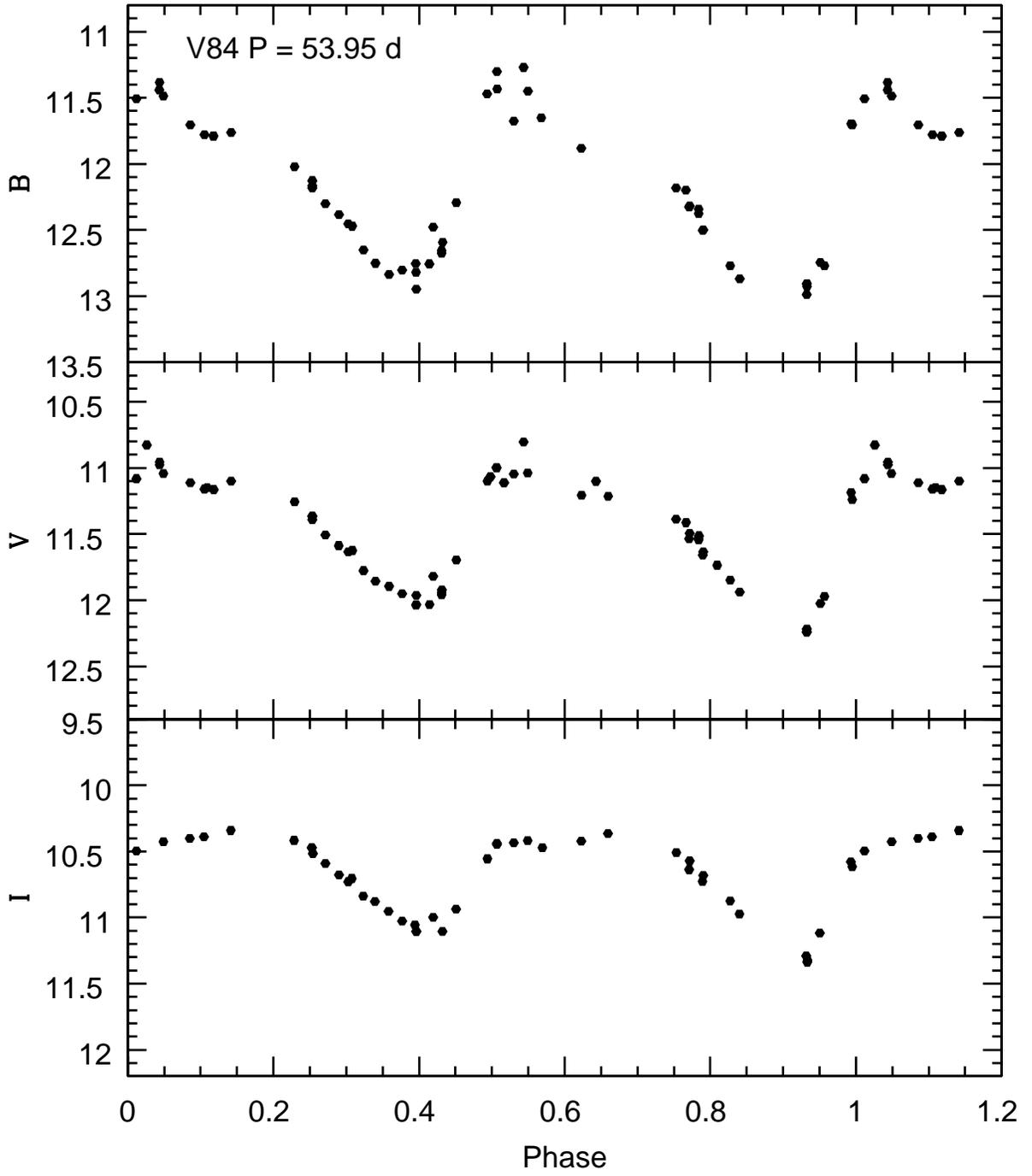}
\caption{Light curves for V84 in M5, phased with a period of 53.95 days.
      }
\end{figure}

\begin{figure}
\includegraphics{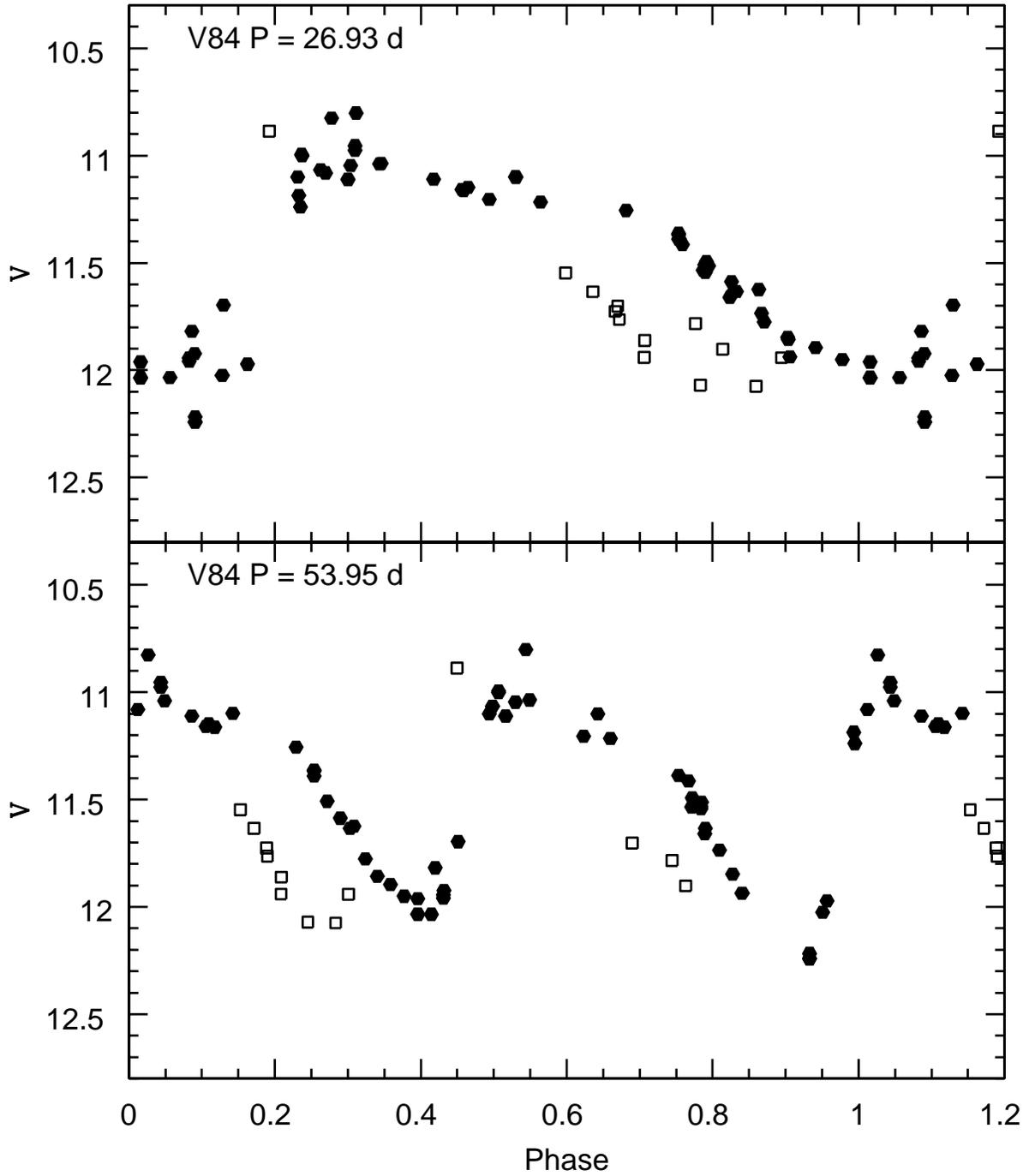}
\caption{Observations from 2006 (open points) and from 2003-2005 (filled points)
for the V light curve of V84.  The light curves in the other pass bands show similar shifts
in phase.
      }
\end{figure}

\begin{figure}
\includegraphics{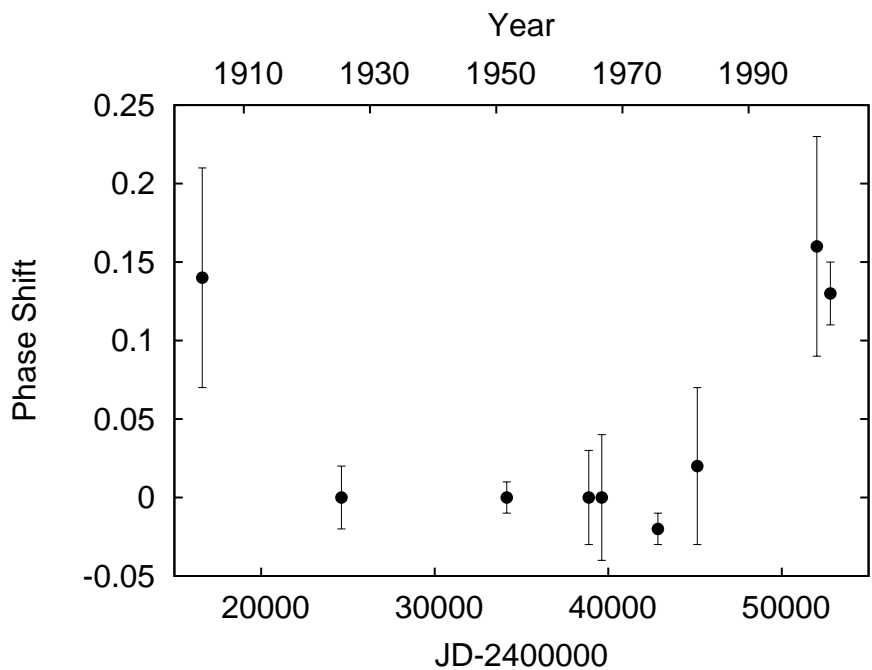}
\caption{Phases of maximum versus Julian Date for V154, calculated assuming a
constant period of 15.2854 days.
      }
\end{figure}

\begin{figure}
\includegraphics{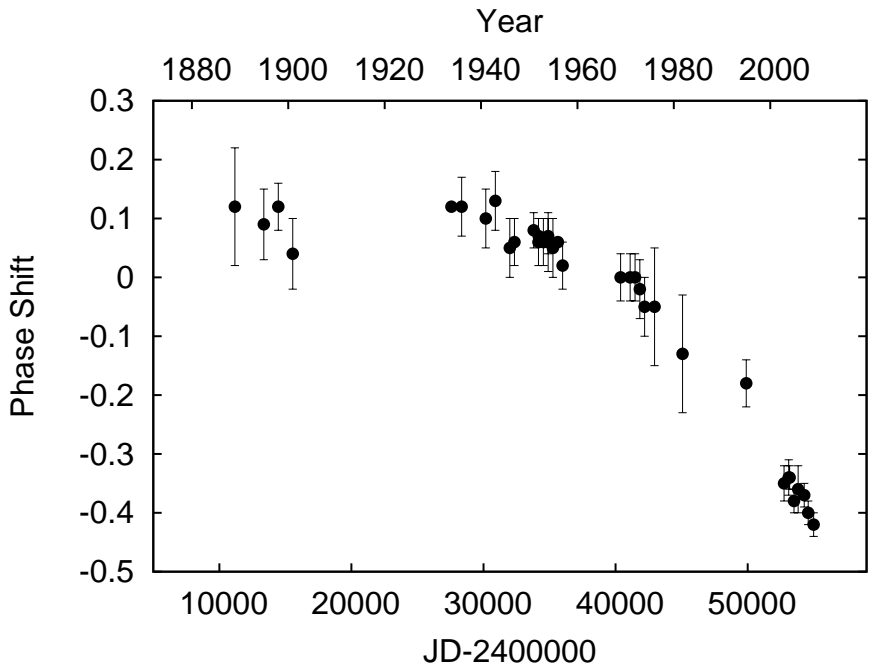}
\caption{Phase shift versus Julian Date for V42, calculated assuming a
constant period of 25.738 days. As indicated in Table V of \citet{cc77},
uncertainties are unknown for a couple of the points.
      }
\end{figure}

\begin{figure}
\includegraphics{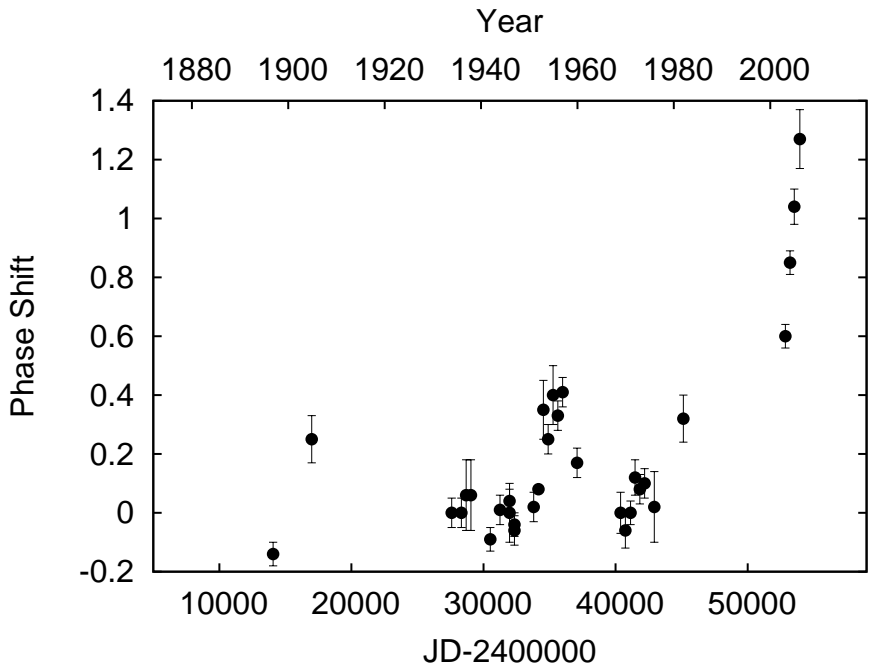}
\caption{Phase shift versus Julian Date for V84, calculated assuming a
constant period of 26.42~d. Error bars are not available for one of the 
points, as indicated in Table V of \citet{cc77}.
      }
\end{figure}

\begin{figure}
\includegraphics{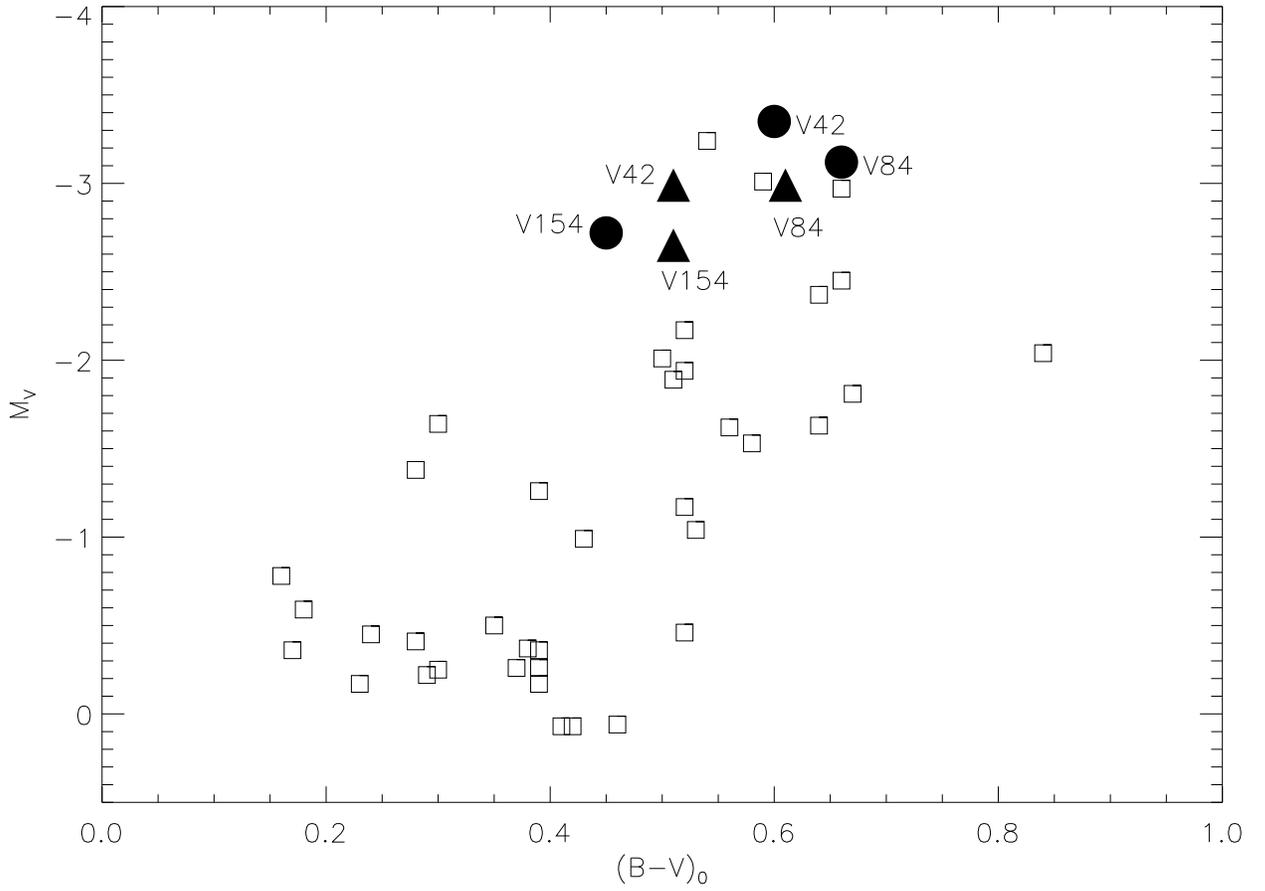}
\caption{Color-magnitude diagram showing type II Cepheids.  Open squares and filled triangles represent
variables in \citet{n94}. Filled circles represent our results for V154, V42, and V84.
      }
\end{figure}

\begin{deluxetable}{rrrrrr} 
\tablecolumns{6} 
\tablewidth{0pc} 
\tablecaption{$BVI_c$ Photometry of V154 in M3, V42 in M5, and V84 in M5}
\tablehead{ 
\colhead{Filter}    &  \colhead{ID} & \colhead{HJD}   &  \colhead{Mag}  & \colhead{Error} & \colhead{Source}}
\startdata 
B	& V154 & 2452763.5450	& 13.24 & 0.03 & MSU \\
B	& V154 & 2452763.5474	& 13.26	& 0.03 & MSU \\
B	& V154 & 2452782.5842	& 13.21	& 0.02 & MSU \\
\enddata
\tablecomments{The full version of this table is included in the electronic form of the paper.}
\end{deluxetable}

\begin{deluxetable}{lll}
\tablecolumns{3}
\tablewidth{0pc}
\tablecaption{Yerkes Observations}
\tablehead{
\colhead{HJD}  & \colhead{B(V42)} & \colhead{B(V84)}  }
\startdata
2445051.9258 &	12.35 & 12.0 \\
2445051.9335 &	12.45 & 11.95 \\
2445057.8720 & 12.5  & 12.2 \\
2445057.8817 &	12.85 & 12.35 \\
2445073.8175 & 11.7 & 11.5 \\
2445073.8266 &	11.95 & 11.6 \\
2445122.7101 &	11.4 & 12.35 \\
2445122.7178 &	11.75 & 12.45 \\
2445128.6939 &	11.75 & 11.45 \\
2445128.7029 &	12.2 & 11.7 \\
2445162.6997 &	12.65 & 11.85 \\
2445162.7080 &	12.35 & 11.95 \\
2445492.6318 &	12.35 & 12.45 \\
2445492.6367 &	12.55 & 12.5 \\
\enddata
\end{deluxetable}

\begin{deluxetable}{llll}
\tablecolumns{4}
\tablewidth{0pc}
\tablecaption{Phases of Maximum for V154 in M3}
\tablehead{
\colhead{JD}  & \colhead{Phase} & \colhead{Uncertainty}  &
\colhead{Source}}
\startdata
2416604.6 & 0.14 & 0.07 & \citet{b06} \\
2424627.55 & 0.00 & 0.02 &  \citet{g35} \\
2434150.4 & 0.00 & 0.01 & \citet{a55} \\
2438873.53 & 0.00 & 0.03 & \citet{k72} \\
2439622.55 & 0.00 & 0.04 & \citet{me80} \\
2442862.71 & -0.02 & 0.01 & \citet{h80} \\
2451256  & 0.08 & 0.08 & \citet{be06} \\
2452021.5  & 0.16 & 0.07 & \citet{st02}\\
2452800.6 & 0.13 & 0.02 & This paper \\
\enddata
\end{deluxetable}

\begin{deluxetable}{llll}
\tablecolumns{4}
\tablewidth{0pc}
\tablecaption{Phase Shifts for V42 in M5}
\tablehead{
\colhead{JD}  & \colhead{Phase} & \colhead{Uncertainty}  &
\colhead{Source}}
\startdata
2445073	& -0.13	& 0.1	& This paper \\
2449890	& -0.18	& 0.04	& Corwin (priv comm.)\\
2453170	& -0.34	& 0.02	& This paper \\
2452750	 &-0.35	& 0.03	& This paper \\
2453106	& -0.34	& 0.03	& This paper \\
2453509	& -0.38	& 0.02	& ASAS \\
2453821	& -0.36	& 0.04	& ASAS \\
2454284	& -0.37	& 0.02	& ASAS \\
2454583	& -0.4	& 0.02	& ASAS \\
2454991	& -0.42	& 0.02	& ASAS \\
\enddata
\end{deluxetable}

\begin{deluxetable}{lll}
\tablecolumns{3}
\tablewidth{0pc}
\tablecaption{Phase Shifts for V84 in M5 from our Observations}
\tablehead{
\colhead{JD}  & \colhead{Phase} & \colhead{Uncertainty} }
\startdata
2445128	& 0.32	& 0.08 \\
2452850 & 0.60 & 0.04 \\
2453200 & 0.85 & 0.04  \\
2453522 & 1.04 & 0.06  \\
2453951 & 1.27 & 0.10 \\
\enddata
\end{deluxetable}

\begin{deluxetable}{lcccc}
\tablecolumns{5}
\tablewidth{0pc}
\tablecaption{Mean Magnitudes and Colors for the Type II Cepheids}
  \tablehead{
\colhead{ID} & \colhead{$<V>_{int}$} & \colhead{$<(B-V)>_{mag}$} & \colhead{$<(V-I)>_{mag}$} &
 \colhead{$<M_V>$} }
\startdata
V154 & 12.33 & 0.46 & 0.65 & -2.72 \\
V42 & 11.19 & 0.63 & 0.76 & -3.35 \\
V84 & 11.42 & 0.69 & 0.80 & -3.12 \\
\enddata
\end{deluxetable}


\begin{thebibliography}

\bibitem[Arp(1955)]{a55} Arp, H.~C.\ 1955, \aj, 60, 1 

\bibitem[Arp(1957)]{a57} Arp, H.~C.\ 1957, \aj, 62, 129 

\bibitem[Arp(1962)]{a62} Arp, H.\ 1962, \apj, 135, 311 

\bibitem[Baade(1956)]{b56} Baade, W.\ 1956, \pasp, 68, 5 

\bibitem[Bakos et al.(2000)]{ba00} Bakos, G.~A., Benko, 
J.~M., \& Jurcsik, J.\ 2000, Acta Astronomica, 50, 221 

\bibitem[Barnard(1898)]{b1898}Barnard, E.~E.\ 1898, 
Astronomische Nachrichten, 147, 243 

\bibitem[Barnard(1906)]{b06}  Barnard, E.~E.\ 1906, 
Astronomische Nachrichten, 172, 345 

\bibitem[Benk{\H o} et al.(2006)]{be06}Benk{\H o}, J.~M., 
Bakos, G.~{\'A}., \& Nuspl, J.\ 2006, \mnras, 372, 1657 

\bibitem[Berdnikov et al.(2009)]{be09}Berdnikov, L.~N.,Henden, A.~A., Turner, D.~G., 
\& Pastukhova, E.~N.\ 2009, Astronomy Letters, 35, 406 

\bibitem[Bidelman(1990)]{b90} Bidelman, W.~P.\ 1990, Information Bulletin on Variable Stars, 3543, 1 

\bibitem[Bono et al.(1997)]{bo97}Bono, G., Caputo, F., \& Santolamazza, P.\ 1997, \aap, 317, 171 

\bibitem[Burwell et al.(1995)]{b95}Burwell, T.~C., Corwin, 
T.~M., Carney, B.~W., Latham, D.~W., \& Danford, S.~C.\ 1995, Bulletin of the American Astronomical Society, 27, 1429 

\bibitem[Cacciari \& Clementini(2003)]{ca03}Cacciari, C., \& Clementini, G.\ 2003, Stellar Candles for the Extragalactic Distance Scale, 635, 105 

\bibitem[Cacciari et al.(2005)]{ca05}Cacciari, C., Corwin, 
T.~M., \& Carney, B.~W.\ 2005, \aj, 129, 267 

\bibitem[Carney et al.(1998)]{car98} Carney, B.~W., Fry, 
A.~M., \& Gonzalez, G.\ 1998, \aj, 116, 2984 

\bibitem[Clement et al.(1988)]{cl88} Clement, C.~M., Hogg, 
H.~S., \& Yee, A.\ 1988, \aj, 96, 1642 


\bibitem[Coutts Clement \& Sawyer Hogg(1977)]{cc77} Coutts Clement, C.~M., \& Sawyer Hogg, H.\ 1977, \jrasc, 71, 281 

\bibitem[Eddington(1918)]{ed18} Eddington, A.~S.\ 1918, 
\mnras, 79, 2 

\bibitem[Gillet(1992)]{g92} Gillet, D.\ 1992, \aap, 259, 215 

\bibitem[Gingold(1976)]{g76}Gingold, R.~A.\ 1976, \apj, 
204, 116 

\bibitem[Gingold(1985)]{g85}Gingold, R.~A.\ 1985, Memorie 
della Societa Astronomica Italiana, 56, 169 

\bibitem[Greenstein(1935)]{g35}Greenstein, J.~L.\ 1935, 
Harvard College Observatory Bulletin, 901, 11 

\bibitem[Hopp(1980)]{h80} Hopp, U.\ 1980, Information 
Bulletin on Variable Stars, 1857, 1 

\bibitem[Kholopov(1972)]{k72}Kholopov, P.~N.\ 1972, 
Astronomicheskij Tsirkulyar, 676, 7 

\bibitem[Landolt(1992)]{la92}Landolt, A.~U.\ 1992, \aj, 
104, 340 

\bibitem[Lenz(2004)]{l05}Lenz, P.\ 2004, Communications in 
Asteroseismology, 144, 41 

\bibitem[Meinunger(1980)]{me80} Meinunger, I.\ 1980, 
Zentralinstitut fuer Astrophysik Sternwarte Sonneberg Mitteilungen ueber 
Veraenderliche Sterne, 8, 161 

\bibitem[Mengel(1973)]{m73} Mengel, J.~G.\ 1973, IAU 
Colloq.~21: Variable Stars in Globular Clusters and in Related Systems, 36, 
214 

\bibitem[Nemec et al.(1994)]{n94}Nemec, J.~M., Nemec, 
A.~F.~L., \& Lutz, T.~E.\ 1994, \aj, 108, 222 


\bibitem[Packer(1890)]{p90} Packer, D.E., 1890, English Mechanic, 51, 378

\bibitem[Percy \& Coffey(2005)]{p05} Percy, J.~R., \& Coffey, J.\ 2005, Journal of the American Association of Variable Star Observers (JAAVSO), 33, 193 


\bibitem[Pickering(1889)]{p89} Pickering, E.~C.\ 1889, 
Astronomische Nachrichten, 123, 207 

\bibitem[Pojmanski(2002)]{po02} Pojmanski, G. 2002, Acta Astronomica, 52,397

\bibitem[Reid(1996)]{r96} Reid, N.\ 1996, \mnras, 282, 304 

\bibitem[Schild(1983)]{sc83} Schild, R.~E.\ 1983, \pasp, 95, 
1021 

\bibitem[Schild(1985)]{sc85} Schild, R.\ 1985, \pasp, 97, 
824 

\bibitem[Schwarzschild \& Harm(1970)]{sch70} Schwarzschild, M., Harm, R.\ 1970, \apj, 160, 341 

\bibitem[Shapley(1914)]{sh14} Shapley, H.\ 1914, \apj, 40, 
448 

\bibitem[Stellingwerf(1978)]{st78} Stellingwerf, R.~F.\ 
1978, \apj, 224, 953 

\bibitem[Sterken(2005)]{st05}Sterken, C.\ 2005, The 
Light-Time Effect in Astrophysics: Causes and cures of the O-C diagram, 
335, 3 

\bibitem[Stetson(1987)]{st87}Stetson, P.~B.\ 1987, \pasp, 
99, 191 

\bibitem[Stetson(1994)]{st94} Stetson, P.~B.\ 1994, 
Astronomy with the CFHT Adaptive Optics Bonnette, 72 

\bibitem[Stetson(2000)]{st00} Stetson, P.~B.\ 2000, \pasp, 
112, 925 

\bibitem[Storm et al.(1991)]{st01} Storm, J., Carney, B.~W., 
\& Beck, J.~A.\ 1991, \pasp, 103, 1264 

\bibitem[Strader et al.(2002)]{st02}Strader, J., Everitt, 
H.~O., \& Danford, S.\ 2002, \mnras, 335, 621 

\bibitem[Templeton \& Henden(2007)]{th07} Templeton, M.~R., \& Henden, A.~A.\ 2007, \aj, 134, 1999 

\bibitem[Turner et al.(2009)]{tu09}Turner, D.~G., Percy, J.~R., Colivas, T., Berdnikov, L.~N., 
\& Abdel-Latif, M.~A.-S.\ 2009, American Institute of Physics Conference Series, 1170, 167

\bibitem[Wallerstein(1958)]{w58}Wallerstein, G.\ 1958, 
\apj, 127, 583 

\bibitem[Wallerstein \& Cox(1984)]{wc84} Wallerstein, G., \& Cox, A.~N.\ 1984, \pasp, 96, 677 

\bibitem[Yong et al.(2008)]{y08}Yong, D., Lambert, D.~L., 
Paulson, D.~B., \& Carney, B.~W.\ 2008, \apj, 673, 854 

\bibitem[Zinn \& West(1984)]{zw84}Zinn, R., \& West, M.~J.\ 1984, \apjs, 55, 45 


\end{thebibliography}
\end{document}